\documentclass[english,preprint,JIP]{ipsj}

\usepackage{graphicx}
\usepackage{latexsym}
\usepackage[hyphens]{url}

\def\Underline{\setbox0\hbox\bgroup\let\\\endUnderline}
\def\endUnderline{\vphantom{y}\egroup\smash{\underline{\box0}}\\}
\def\|{\verb|}

\usepackage[varg]{txfonts}
\makeatletter%
\input{ot1txtt.fd}
\makeatother%

\begin{document}

\title{Zero Trust Federation: Sharing Context under\\User Control toward Zero Trust in Identity Federation}

\affiliate{KU}{Kyoto University}

\author{Koudai Hatakeyama}{KU}[hatakeyama@net.ist.i.kyoto-u.ac.jp]
\author{Daisuke Kotani}{KU}[kotani@media.kyoto-u.ac.jp]
\author{Yasuo Okabe}{KU}[okabe@media.kyoto-u.ac.jp]

\begin{abstract}
To securely control access to systems, the concept of Zero Trust has been proposed.
Access Control based on Zero Trust concept removes implicit trust and instead focuses on evaluating trustworthiness at every access request by using contexts.
Contexts are information about the entity making an access request like the user and the device status.
Consider the scenario of Zero Trust in an identity federation where the entity (Relying Party; RP) enforces access control based on Zero Trust concept.
RPs should continuously evaluate trustworthiness by using collected contexts by themselves, but RPs where users rarely access cannot collect enough contexts on their own.
Therefore, we propose a new federation called Zero Trust Federation (ZTF).
In ZTF, contexts as well as identity are shared so that RPs can enforce access control based on Zero Trust concept.
Federated contexts are managed by a new entity called Context Attribute Provider, which is independent of Identity Providers.
We design a mechanism sharing contexts among entities in a ZTF by using the two protocols; context transport protocol based on Continuous Access Evaluation Protocol and user consent protocol based on User Managed Access.
We implemented the ZTF prototype and evaluated the capability of ZTF in 4 use-cases.
\end{abstract}

\begin{ekeyword}
Zero Trust, Identity Federation
\end{ekeyword}

\maketitle

\section{Introduction}
To securely restrict access to systems and resources, organizations need access control to determine who can gain access to what.
When organizations try to protect their systems against the sophisticated attackers that have emerged in recent years, it is preferable to enforce access control based on Zero Trust.
Zero Trust, which is proposed by John Kindervag \cite{forrester-whitepaper}, is the term for describing various cybersecurity solutions that remove implicit trust and instead focus on evaluating trustworthiness at every access request \cite{nist800-207}.

For evaluating trustworthiness at every access request, identity and contexts are needed.
(Digital) Identity is information to identify the entity making an access request.
Contexts are various information about the entity making an access request: information about the user, the device being employed by her, the network to which her device is connected, the physical environment surrounding her, etc.
Contexts include not only static information as device vendors and user roles but also dynamic information based on past behavior, such as what device was used for recent access.
Since organizations can evaluate how trustworthiness she proves by using contexts, they can determine whether she is enough trustworthy to do what she wants to do.

Consider now that the entity which controls access to its services based on Zero Trust joins an identity federation.
Since the entity (Relying Party; RP) belongs to the identity federation, the RP can receive federated identity from the Identity Provider (IdP) that authenticates users.
However, identity is not sufficient for the RP to evaluate the trustworthiness of every access request.
The IdP can verify an access request to the RP only when the IdP tries to authenticate the user making the access request.
Therefore, the RP should continuously collect contexts by themselves and evaluate the trustworthiness by using the collected contexts as well as federated identity.

However, in an identity federation, some RP cannot have enough contexts on their own, because contexts included in access requests are collected by RPs only when users access themselves.
The RPs where users rarely access store little contexts on their own, but identity federations cannot share contexts like identity.

To share contexts in an identity federation, Continuous Access Evaluation Protocol \cite{caep-blog} has been proposed.
In this protocol, IdPs collect contexts not only on their own but also from RPs federating with the IdPs.
IdPs can provide federated contexts to the RPs that want to receive the contexts.
Here federated contexts mean the contexts which are shared among multiple entities.

Although IdPs can probably have the capabilities to share federated contexts which they collect if an identity federation has that capability, only IdPs shouldn't be the entity that manages federated contexts because of the following three reasons.

The first reason is that the lifecycle of contexts is not always the same lifecycle of identity managed by IdPs.
For example, when your university IdP manages the context about what device you have used for recent access and you often use your laptop, after graduating you can no longer use the context even though you can still use the same laptop.
Contexts should not be bound to identity managed by an IdP within an identity federation. 
Instead, contexts should be associated with the entity that is the subject of the contexts across identity federations.

The second reason is that some contexts are collected and managed by a designated entity that is independent of IdP and RP.
Endpoint Detection and Response service is an example of this entity. 
This service collects and manages device health status such as whether devices are compromised through the agent software installed in each device.

The third reason is that RPs should mitigate implicit trust derived from a single IdP.
When a single IdP provides contexts as well as identity to RPs, RPs implicitly rely on the trust derived from the single IdP to enforce access control by using the federated contexts and identity.
Once an IdP is compromised or cannot handle credential leaks properly, the attacker can impersonate a legitimate user using her federated identity issued by the IdP.
By using federated contexts from various entities other than an IdP, RPs should minimize implicit trust of the IdP when evaluating the trustworthiness of an access request.

Therefore, we propose a new federation called Zero Trust Federation (ZTF).
In ZTF, federated contexts, as well as federated identity, are shared among the entities so that RPs can enforce access control based on Zero Trust by using these contexts to evaluate the trustworthiness at every access request.
We introduce a new entity called Context Attribute Provider into ZTF which collects, manages, and shares federated contexts and which is independent of IdP and RP.

Furthermore, we note that contexts are privacy sensitive information.
Sharing federated contexts must be under user control.
Therefore, we designed the mechanism in which user authorization is mandatory to establish sharing contexts.

We designed a mechanism that shares federated contexts among systems operated by entities joining a ZTF.
The mechanism consists of two protocols; a transport protocol that allows the entity in a ZTF to convey collected contexts to others and a user consent protocol that allows users who are subjects of contexts to grant access to contexts only to authorized entities by the user.
The former protocol is based on a protocol for security-event sharing called Continuous Access Evaluation Protocol, which is currently being standardized by Shared Signals and Events WG of OpenID Foundation.
The latter protocol is based on a protocol for access control management called User Managed Access, which extends an authorization delegation protocol called OAuth2.0.
Furthermore, we implemented a prototype complying with the mechanism for evaluation.

The main contribution of this study is that it proposes the concept of Zero Trust Federation (ZTF) that extends identity federation by introducing a new entity called Context Attributed Provider that collects, manages, and shares federated contexts.
ZTF allows RPs to enforce access control based on Zero Trust by using federated contexts and federated identity to evaluate the trustworthiness of every access request.

The following is the structure of this paper.
Section 2 describes the background, and Section 3 describes the concept of Zero Trust Federation (ZTF). Section 4 describes user authorization for sharing contexts.
Section 5 describes evaluation with use cases using the prototype that implements ZTF.
Finally, Section 6 summarizes this paper.

\section{Background}
\subsection{Zero Trust Network (ZTN)}
ZTN is an access control model whose core principle is ``Never Trust, Always Verify'' \cite{forrester-whitepaper}.
ZTN does not rely on a single, implicit trust for access control like source networks in traditional perimeter models (``Never Trust'').
Instead, it verifies every access request by using contexts.
Contexts are information about the entity making an access request: information about the user, the device being employed by her, the network to which her device is connected, the physical environment surrounding her, etc.
The context includes both static information as user IDs and device vendors and dynamic information based on past behavior, such as what device was used for recent access and where it was accessed.

According to Gilman \cite{oreilly} and NIST SP 800-207 \cite{nist800-207}, Zero Trust Architecture (ZTA) consists of the control plane and data planes.
The control plane, which must determine whether to allow access to protected resources, is called a Policy Decision Point (PDP) \cite{nist800-207}.
Data planes are the place where a user communicates with a resource and where access control is enforced.
An access control enforcement point is also called a Policy Enforcement Point (PEP) \cite{nist800-207}.

BeyondCorp is Google's implementation of ZTN in its internal network \cite{beyondcorp43231}\cite{beyondcorp44860}\cite{beyondcorp45728}.
This implements an access proxy that combines PEP and PDP to enable ZTN access control.
This (reverse) proxy, also known as an Identity-Aware Proxy (IAP) \footnote{\url{https://cloud.google.com/iap/}}, collects contexts, verifies access requests, and enforces authorization decisions. 
By combining PEP and PDP, policy changes can be applied quickly and consistently.
However, it is difficult that one IAP protects all RPs in an identity federation to collect enough contexts from access requests to these RPs.
This is because it is impossible that one IAP understands and verifies all access requests to each RPs that provide various services.
Even if the IAP can do that, RPs rely on implicit trust derived from the IAP.

\subsection{Contexts Used for Verification}
ZTN uses contexts to verify every access request to authenticate the user who requested access and to authorize it.

Context-based authentication and authorization (AuthNZ), also known as risk-based AuthNZ, uses various contexts.
These contexts include IP addresses \cite{ctx-authn-diep2007contextual}, mouse movements and keystrokes \cite{ctx-authn-6376399}, touchscreen input \cite{ctx-authn-6331527}, fingerprinting of various user information \cite{ctx-authn-10.1145/2695664.2695908}, physical location of users \cite{ctx-authz-MINAMI2005123}, environments surrounding devices \cite{ctx-authz-10.1145/373256.373258}, etc.

As the above examples of contexts, contexts can be collected only by the entities that users directly access and the entities that can let users install the agent software in each device.
Therefore, not all the entities in identity federations can utilize such contexts.

\subsection{Identity Federation (IdF)}
IdF is a group of organizations that agree to follow the rules of a trust framework.
A trust framework is the rules underpinning federated identity management, typically consisting of: system, legal, conformance, and recognition \cite{nistir8149}.
Here federated identity is the information that an IdP issues and that a RP uses to identify a user.
Since an IdP signs federated identity, a RP can verify the integrity of the federated identity and whether the federated identity is issued from the IdP that the RP relies on.
IdPs and RPs joining an IdF agree not only on technical rules such as the transport protocol and identity representation but also on contract to define responsibility for IdPs to manage identity. 

An example of an inter-university IdF is Gakunin \footnote{\url{https://www.gakunin.jp/}} through which users can seamlessly access services provided by different universities without having to register as a new user, for example, connecting to wireless LANs at different universities without registration.

Shibboleth \footnote{\url{https://www.shibboleth.net/}} using SAML\cite{saml-v2} or OpenID Connect \cite{oidc-core} are examples of protocol that establish IdFs.

\subsection{Continuous Access Evaluation Protocol (CAEP)}
CAEP is a new event-sharing protocol for Continuous Authentication in an IdF \cite{caep-blog}.
With CAEP, RPs can share internally generated events about users with the IdP that authenticates them.
Events include a change of the network being used or a vulnerability that is discovered in the device being used.
In other words, an IdP can sense updates for user contexts that occurred at RPs with this protocol and manage continuous authentication based on these notified contexts.
In addition, an IdP can notify RPs of the authentication results and such additional information as events about Authenticator Assurance Level \footnote{\url{https://pages.nist.gov/800-63-3/sp800-63b.html}} changes due to stronger authentication for other RPs.
RPs make authorization decisions using contexts about the user based on this information.

This protocol is currently being standardized in the Shared Signals and Events Working Group in the OpenID Foundation\footnote{\url{https://openid.net/wg/sse/}}.

\subsection{User Control in Federation}
Sharing privacy sensitive information must be under user control.
For example, Gakunin shares attributes of users as well as their pseudonymous IDs among organizations.
With uApproveJP \cite{orawiwattanakul2011user}, user attributes can be shared with a user's consent in the identity federation.

OAuth2.0 \cite{rfc6749}, which is a protocol for authorization delegation, allows a third party (client) to access a resource server on behalf of the resource owner who provides permission within a limited access scope.
On a resource server, resource owners deploy their owned resources.
Resources are protected by an authorization server, which asks resource owners for an authorization decision for resource access requests from clients. 
When resource owners approve, the authorization server issues an access token to clients, which represents the granted permissions.
The resource server verifies the presented token and determines the validity of the access.
In this way, the resource owner can delegate authorization to the client with OAuth2.0.

User Managed Access (UMA) is a protocol based on OAuth2.0.
It extends OAuth2.0 from the following two points.
In OAuth2.0, the client is operated by a resource owner who can make authorization decisions. 
But in UMA, the client is operated by an entity called a Requesting Party that requests access to resources, which is not always a resource owner.
For this reason, UMA extends the authorization server to make authorization decisions depending on the policy set by the resource owner in advance.
These extensions allow a Requesting Party to access the resource server within the range defined by the resource owner in its policy.

\section{Zero Trust Federation (ZTF)}
This section defines Zero Trust Federation and proposes the protocol for sharing federated contexts.

\subsection{Requirement}
As mentioned in the introduction, there are problems when RPs enforce access control based on Zero Trust in identity federations.

Therefore, the requirements for new federations are following.
\begin{enumerate}
  \item RPs can collect enough contexts by sharing contexts scattered among RPs and by receiving contexts from the independent entity of IdPs and RPs.
  \item The lifecycle of contexts is not bound to the lifecycle of identity managed by a specific IdP
  \item RPs can use contexts to detect the abnormal behavior of the user who is identified by identity from an IdP
\end{enumerate}

\subsection{Definition}
We define Zero Trust Federation (ZTF) as a federation that allows each Relying Party (RP) to enforce access control based on Zero Trust by federating with Identity Providers (IdPs) and Context Attribute Providers (CAPs).
IdPs identify and authenticate users and provide federated identity to other entities.
CAPs collect and manage contexts and provide federated contexts to other entities.
Federated contexts mean the contexts shared in multiple entities.
RPs, which users request access to, use federated identity from IdPs and federated contexts from CAPs to evaluate the trustworthiness of each access request for enforcing access control based on Zero Trust. 

\begin{figure}[htb]
  \centering
  \includegraphics[width=0.45\textwidth]{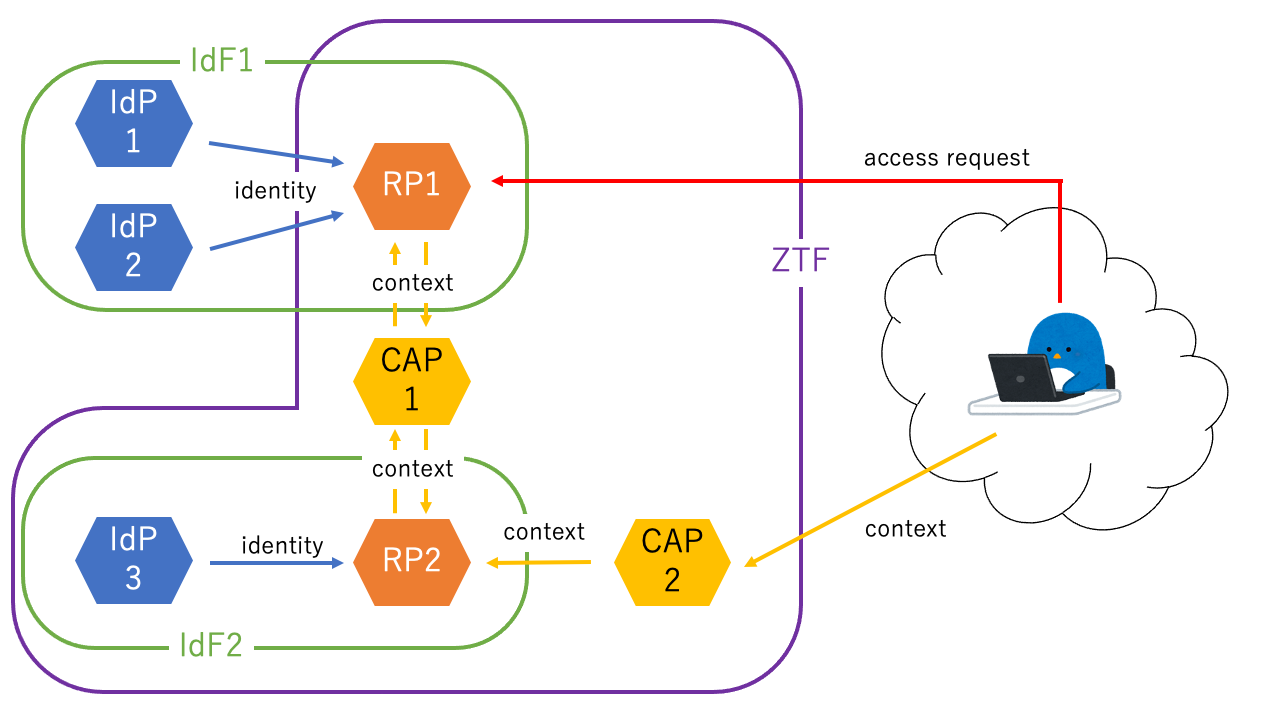}
  \caption{ZTF Overview}
  \label{fig:ztf:overview}
\end{figure}

As shown in Fig.\ \ref{fig:ztf:overview}, ZTF meets the above requirements.
First, CAP1 collects contexts from RP1 and RP2 and provides federated contexts to these RPs.
Since CAP1 makes these federated contexts by combining the contexts from RP1 with the contexts from RP2, RPs can use enough contexts to make authorization decisions.
Besides, RP2 can receive contexts from CAP2 that directly collects contexts from users and that is an independent entity of IdPs and RPs.

Second, since CAPs are independent of IdPs, even if users change IdP1 to IdP2 when they access RP1, i.e., the federated identity is changed, users can still use federated contexts provided by CAP1 when they access RP1 as usual.

Finally, even if IdP3 is compromised and attackers can impersonate legitimate users, RP2 can use federated contexts from CAP1 and CAP2 to detect the suspicious behavior of the user identified by the federated identity from compromised IdP3.

\subsection{Protocol for Sharing Contexts}
In this study, the protocol for sharing contexts is based on CAEP \cite{caep-blog}.
However, since CAEP has not been standardized yet, we design it based on the current draft.

\subsubsection{Representation of Contexts}
Contexts are represented by an extended JWT\footnote{\url{https://tools.ietf.org/html/rfc7519}} called SET\footnote{\url{https://tools.ietf.org/html/rfc8417}}, as proposed in CAEP.
The example of context representation is the following.
This context is provided by CAP1 (\verb|https://cap1.example|) to RP1 (\verb|https://rp1.example|).
It describes whether the IP address (\verb|192.0.2.1|) currently used by the laptop operated by Alice has been already used by the same laptop in the past.

A CAP can set federated contexts to \verb|events| property, in which the CAP sets the kind of the context to a JSON key and sets the content of the context to a JSON value.
The syntax of the content of the context is agreed among RPs that receive the federated context and the CAP that transmits the federated context.

\begin{figure}[htb]
  \centering
  \begin{verbatim}
    { "aud": "https://rp1.example",
      "iat": 1619696843,
      "iss": "https://cap1.exmaple,
      "jti": "metyakutya-random",
      "events": {
        "https://cap1.exmaple/ctxtype/device-location": {
          "subject": {
            "user"  : { "format": "email", 
                        "email" : "alice@exmaple.com" },
            "device": { "format": "cn",
                        "cn"    : "alice-no-Laptop" }},
          "used:ip:192.0.2.1": true
        }}}
    \end{verbatim}
  \caption{The example of context representation}
  \label{fig:ztf:proto:ctx}
\end{figure}

\subsubsection{Transfer and Receive Contexts}
\begin{figure}[htb]
  \centering
  \includegraphics[width=0.45\textwidth]{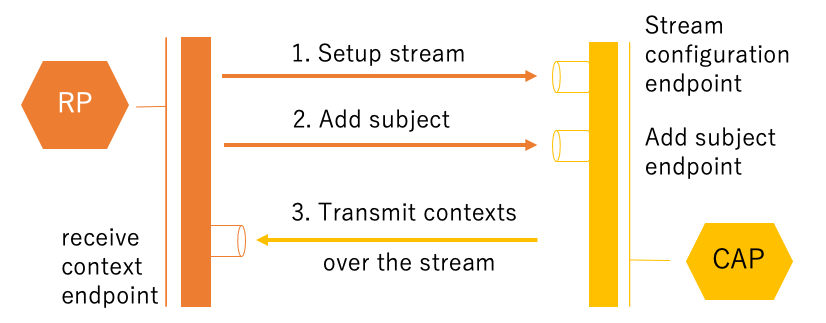}
  \caption{Flow of receiving contexts}
  \label{fig:ztf:proto:caep}
\end{figure}

CAEP is based on an IETF draft called Management API for SET Event Streams\footnote{\url{https://tools.ietf.org/html/draft-scurtescu-secevent-simple-control-plane}}.
In CAEP, contexts are transmitted over a virtual stream.
Consider that some RP wants to receive federated contexts about a user from some CAP.
This stream is set up by the RP sending the setup configuration (Step1 in Fig.\ \ref{fig:ztf:proto:caep}).
The RP requests the CAP to add the user to the stream (Step2).
Once the CAP approves adding the user, the CAP provides the federated contexts about the user over the stream (Step3).
And when the CAP detects a context update, the CAP provides the federated contexts about the user over the stream.

\section{User Authorization for Sharing Contexts}
This section proposes the protocol for federating contexts under user control. 

\subsection{Requirement}
As mentioned in the introduction, there is a privacy problem when sharing contexts, which are privacy sensitive information, among entities in ZTF.

Therefore, the requirements of ZTF for privacy are following.
\begin{enumerate}
  \item When CAPs (RPs) provide federated contexts to RPs (CAPs), users can grant which RPs (CAPs) access to which limited contexts.
  \item Users can set policy about which RPs and CAPs can access what contexts in which scopes without requested to consent each time to establish a new context sharing.
\end{enumerate}

The reason for requirement 2 is not to bother users.
In ZTF, many CAPs provide contexts to many RP and CAPs collect from many RPs so that users have to control many context sharing.

\subsection{User Authorization Service}
\begin{figure}[htb]
  \centering
  \includegraphics[width=0.45\textwidth]{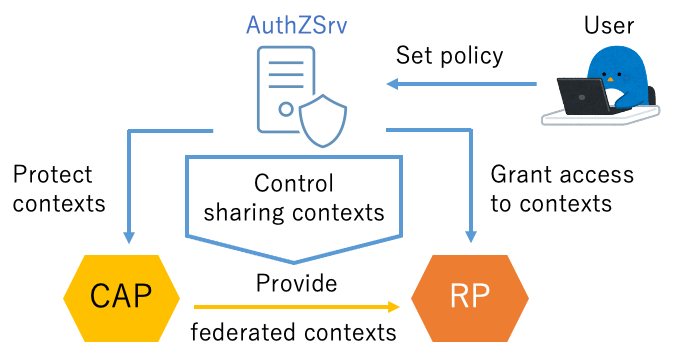}
  \caption{User Authorization Server (AuthZSrv)}
  \label{fig:privacy:uma}
\end{figure}

We introduce a new service for user authorization for sharing contexts into ZTF (Fig.\ \ref{fig:privacy:uma}).
A User Authorization Server (AuthZSrv) makes authorization decisions about which RPs can access which contexts in CAPs.
This decision is based on policies set by users who own contexts.
To protect contexts stored in CAPs, CAPs register the type of contexts to the AuthZSrv on behalf of the user who owns the contexts.
When RPs want to receive federated contexts from CAPs, they request the AuthZSrv to grant access to the contexts.
Since CAPs verify whether the AuthZSrv grants RPs access to federated contexts and enforce the authorization decisions made by the AuthZSrv, users can control sharing contexts in ZTF.

\subsection{Protocol for User Authorization}
In this study, protocol for user authorization is based on User Managed Access \cite{umagrant}\cite{umafed}, which extends OAuth2.0 \cite{rfc6749}.
The following are two advantages of using UMA in ZTF:
1) the AuthZSrv can centrally manage federated contexts in RPs and CAPs, and control which RPs and CAPs can access to what federated contexts; 
2) the AuthZSrv can make authorization decisions automatically based on policies having set up by users without requesting a consent action each time to establish a context sharing with a new RP.

\subsubsection{Registration of Contexts to the AuthZsrv}
Since the AuthZSrv needs to identify contexts that it protects, CAPs register what kind of context they manage to an AuthZSrv.
This registration flow is based on Federated Authorization for UMA \cite{umafed}.

\begin{figure}[htb]
  \centering
  \includegraphics[width=0.45\textwidth]{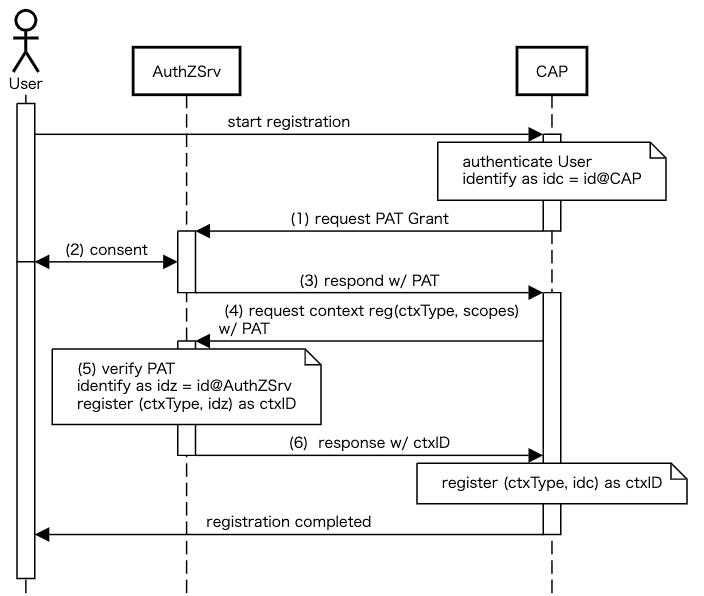}
  \caption{Flow for Context Registration to AuthZSrv}
  \label{fig:privacy:proto:rreg}
\end{figure}

As shown in Fig.\ \ref{fig:privacy:proto:rreg}, the flow for context registration is following.
Before registration, the CAP has to have an access token to register the context owned by the user to the AuthZSrv.
This access token is called Protection API Token (PAT) issued by the AuthZSrv.
At first, (1) The CAP requests the AuthZSrv to grant a PAT of the user.
(2) The AuthZSrv gets consent from the user and (3) responds with the PAT as proof of permission of the user.
Since the CAP gets the PAT to register the context of the user, (4) the CAP requests the AuthZsrv to register the context with presenting the PAT.
(5) The AuthZSrv verifies the PAT, identifies who owns the context in the CAP, and then registers the context by issuing a context identifier.
(6) The AuthZSrv responds with the context identifier so that by using this identifier, the CAP and the AuthZSrv can identify the context of the user.

\subsubsection{Granting RPs access to contexts}
Before RPs request the AuthZSrv to grant access, RPs need to identify contexts owned by the user.
At first, the user registers the context identifier to the RP.
We note that this registration flow is out of scopes of UMA \cite{umagrant}.

\begin{figure}[htb]
  \centering
  \includegraphics[width=0.45\textwidth]{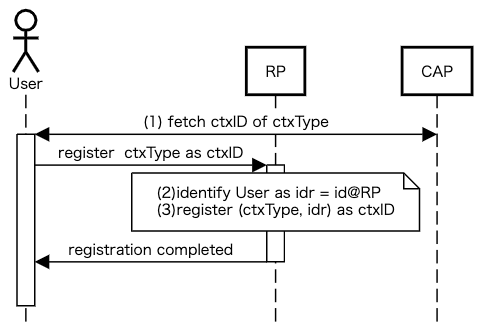}
  \caption{Flow for Context Registration to RP}
  \label{fig:privacy:proto:rreg-to-rp}
\end{figure}

As shown in Fig.\ \ref{fig:privacy:proto:rreg-to-rp}, the flow for context registration to RP is following.
The user fetches the context identifier (ctxID) from the CAP, which refers to the contexts owned by the user and is uniquely identified within the CAP.
RP (2) identifies the user and (3) registers the ctxID of the contexts owned by the user.
The RP and the CAP can then identify the contexts of the user by using this identifier.

\begin{figure}[htb]
  \centering
  \includegraphics[width=0.45\textwidth]{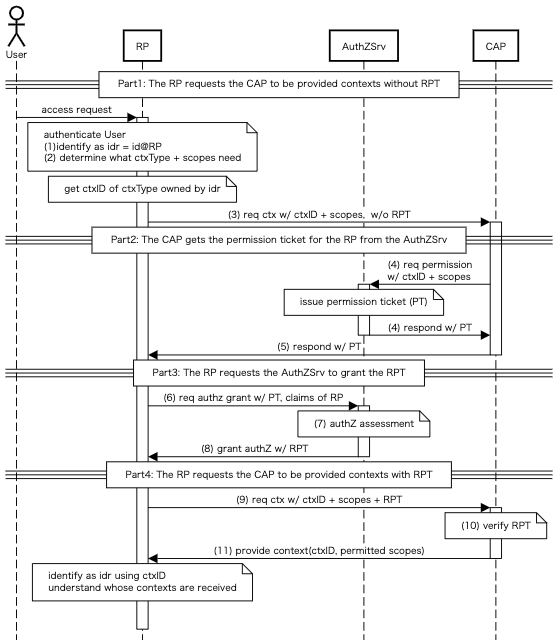}
  \caption{Flow for Granting Authorization to RP}
  \label{fig:privacy:proto:umagrant}
\end{figure}

The granting authorization flow is based on UMA \cite{umagrant}.
As shown in Fig.\ \ref{fig:privacy:proto:umagrant}, the flow for granting authorization is following.
This flow can be divided into four parts.

The first part is that the RP requests the CAP to provide the context without user authorization.
The RP (1) identifies the user making the access request and (2) determines what kind of context (ctxType) needs in which scopes to make the authorization decision.
Since the user has registered the context identifier (ctxID) of the context, (3) the RP requests the CAP to provide contexts with the ctxID and the scopes.

The second part is that the CAP gets a permission ticket from the AuthZSrv.
Since the RP is not approved, the CAP lets the RP request the AuthZSrv to grant user permission.
The CAP (4) gets from the AuthZSrv the permission ticket (PT) which describes what kind of context the RP wants in which scopes, and (5) responds with the PT.

The third part is that the RP requests the AuthZSrv to grant user permission by presenting the PT and its claims.
Since the CAP has provided the PT to the RP, (6) the RP requests the AuthZSrv to grant the permission of the user.
When requesting, the RP also presents the claims about the RP itself.
(7) The AuthZSrv determines whether the user approves the context request from the RP by using the policy having being set.
If approved, (8) the AuthZSrv provides the RP the Requesting Party Token that describes what permissions the RP has been granted.

The final part is that the RP re-requests the CAP to provide the context with the RPT.
(9) The RP requests the CAP to provide contexts with the RPT again.
The CAP (10) verifies the RPT, and (11) provides the contexts of the user limited the permitted scopes.

\section{Implementation Prototype and Use Cases}
We implemented a prototype that satisfies the protocol shown in Sections 3 and 4.

The prototype was created using Golang, with JWT libraries \footnote{\url{https://github.com/lestrrat-go/jwx}} and HTTP utility libraries \footnote{\url{https://github.com/gorilla/}}, and KeyCloak \footnote{\url{https://www.keycloak.org/}} as IdP and Authorization Server.
The source code is available on GitHub \footnote{\url{https://github.com/hatake5051/ztf-prototype/tree/ipsj}}.

\begin{figure}[htb]
  \centering
  \includegraphics[width=0.45\textwidth]{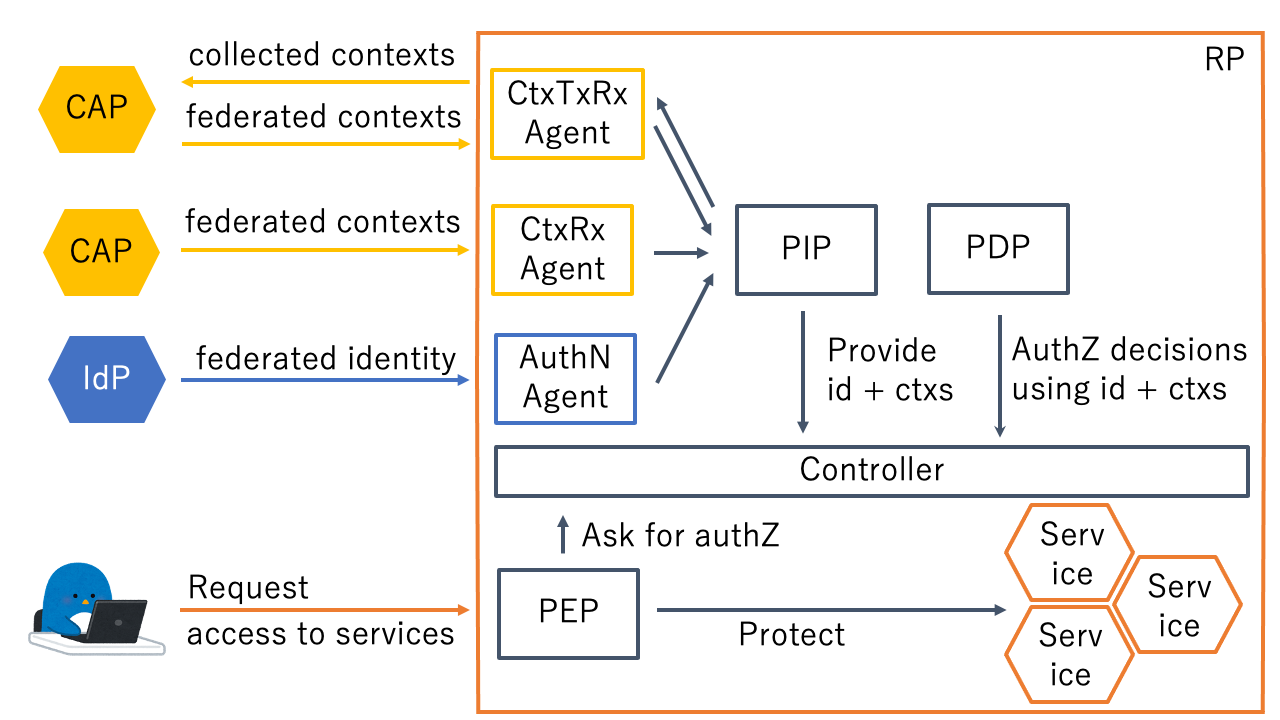}
  \caption{The Design of RP}
  \label{fig:impl:rp}
\end{figure}

Fig.\ \ref{fig:impl:rp} shows an overview of the design of RP.
This design is inspired by XACML Data-flow model \cite{xacml-v3}.
RPs protect their managed services by enforcing access control at Policy Enforcement Point (PEP).
PEP asks Controller for authorization decisions such as whether this access request is approved.
Policy Information Point (PIP) provides identity and contexts to Controller, and Policy Decision Point (PDP) determines authorization decisions.
By using an authentication agent PIP collects federated identity from an IdP.
By using a context receiver agent PIP collects federated contexts from a CAP.
When a CAP collects contexts from a RP, the RP uses a context transmitter agent to provide their collecting contexts to the CAP.

\begin{figure}[htb]
  \centering
  \includegraphics[width=0.45\textwidth]{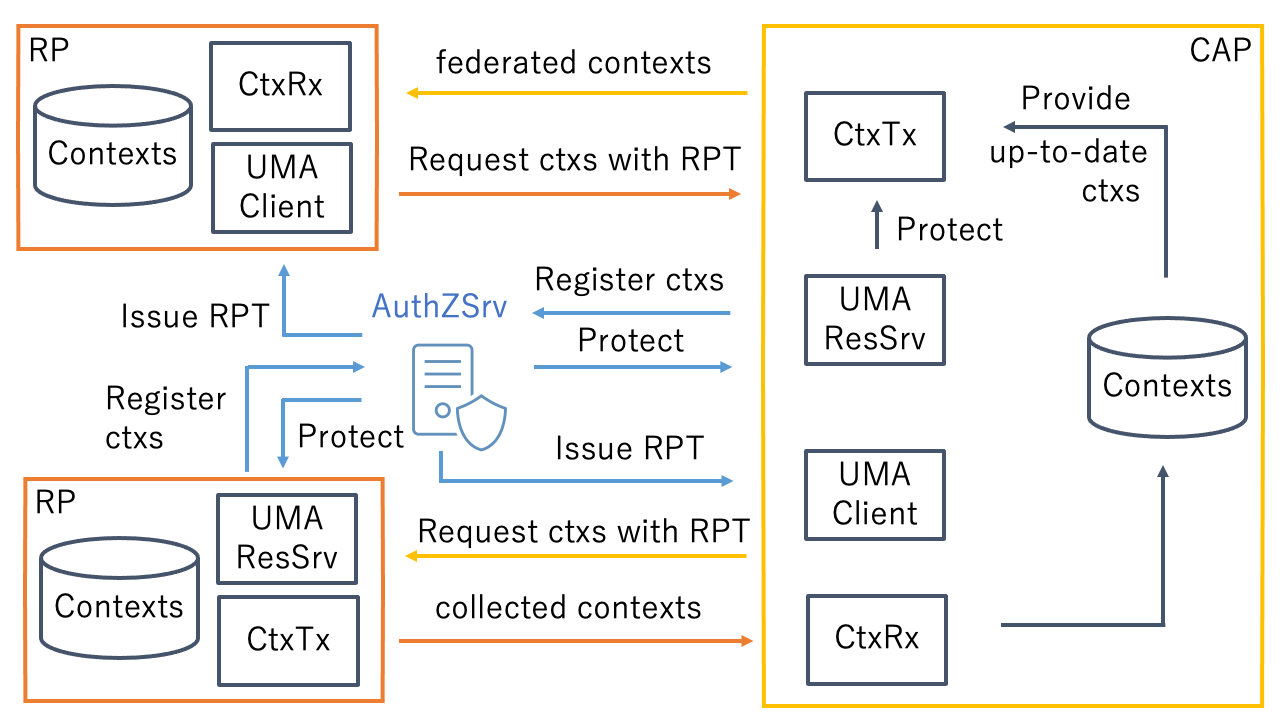}
  \caption{The Design of Federating Contexts under User Control}
  \label{fig:impl:cap}
\end{figure}

Fig.\ \ref{fig:impl:cap} shows an overview of the design of sharing contexts under user control.
When a CAP shares federated contexts with RPs, the CAP uses the UMA Resource Server function to register contexts to the AuthZSrv.
RPs ask the AuthZSrv to grant access to the contexts and, if approved, the AuthZSrv issues Requesting Party Token (RPT) as proof of granted access.
RPs present the RPT to the CAP, and the CAP provides federate contexts using the CAEP Transmitter function.
As vice versa above, when a CAP collects contexts from RPs, the CAP requests RPs to transmit federated context.

\subsection{Evaluation with Use Cases}
\begin{figure}[htb]
  \centering
  \includegraphics[width=0.45\textwidth]{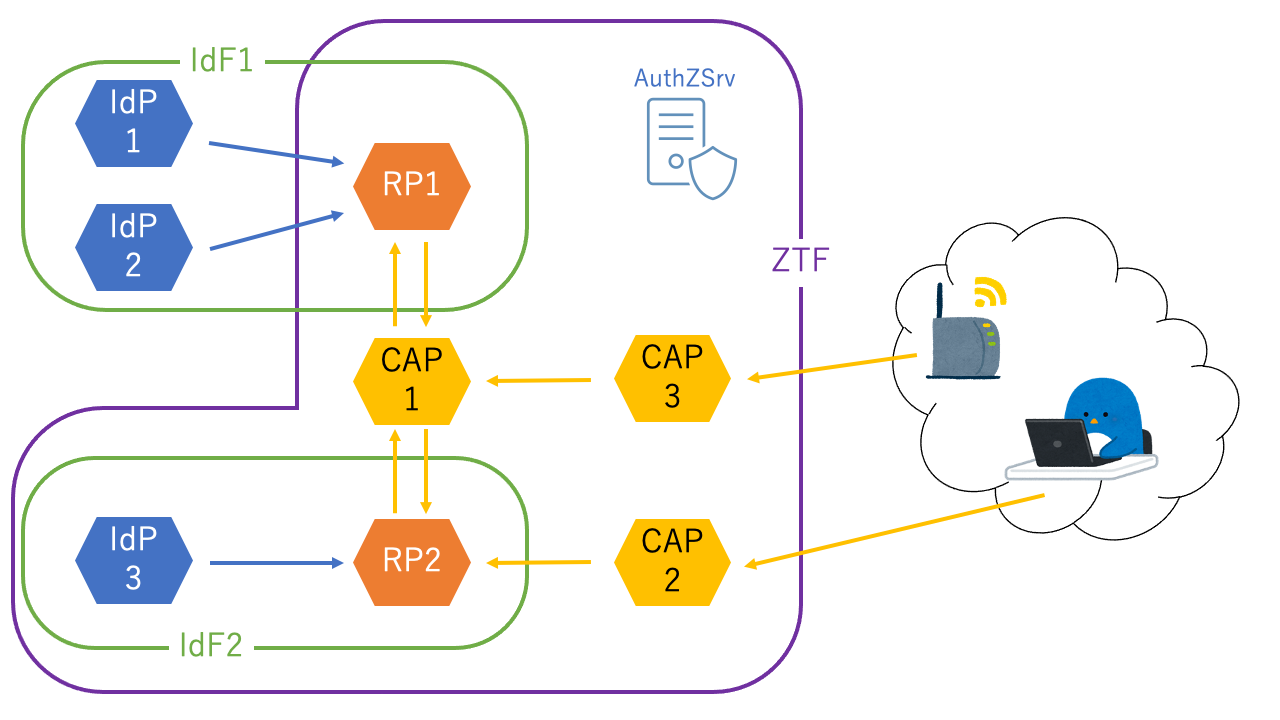}
  \caption{ZTF Use Cases}
  \label{fig:usecases}
\end{figure}

We evaluate ZTF using the following four use cases.
Fig.\ \ref{fig:usecases} shows the ZTF environment used in use cases.
RP1, IdP1, and IdP2 belong to an identity federation called IdF1.
RP2 and IdP3 belong to another identity federation called IdF2.
RP1, RP2, CAP1, CAP2, and CAP3 belong to a ZTF.
Sharing federated contexts under ZTF are controlled by AuthZSrv.

CAP1 collects and manages the contexts about device location.
It manages contexts about which users use what devices and where users use these devices.
It collects device identifiers and network information from RPs.
It provides federated contexts in permitted scopes to RP1 and RP2 when users approve.
For example, the contexts include information about whether the device used in the access request has ever been used by the user and whether the network location used by devices is already used by the device.

CAP2 collects and manages the contexts about device health status.
It manages contexts about which users use what devices and whether the version of these devices is up-to-date.
It collects device identifiers and information about device versions from the agent installed in each device.
It provides federated contexts in permitted scopes to RP2 when users approve.

CAP3 collects and manages the contexts of the physical environment surrounding the user.
It collects contexts from Wi-Fi Access Points and manages contexts about whether the device used in the access request is connected with white-listed Wi-Fi Access Points.
It provides federated contexts to CAP1, and CAP1 aggregates the contexts provided by CAP3 with the contexts collected on their own.

\subsubsection{Providing Contexts from CAP2 to RP2}
Assume that RP2 wants to approve access to its services only from those who use an up-to-date version of devices but RP2 cannot collect those contexts by itself.
Although IdP3 provides federated identity to RP2, IdP3 cannot provide these contexts as well as federated identity because IdP3 does not have the capabilities of collecting device health information.
However, CAP2 can collect device health information through the agent software installed in devices used by users.
In ZTF, CAP2 federates with RP2 and provides these contexts about whether the device used by the user who requested access is an up-to-date version so that RP2 can enforce access control based on its policies.

ZTF allows RPs to receive contexts from the CAP that is independent of IdP.

\subsubsection{Sharing Contexts among RP1 and RP2 via CAP1}
Assume that RP2 and RP1 want to approve access only from the network which is often used by users but RP1 runs services that users use infrequently, such as few times a year.
Maybe RP1 can collect these contexts from IdP1 or IdP2 in IdF1, but IdPs in IdF1 cannot collect enough contexts used for evaluating the frequency since RP2 does not belong to the same identity federation.
However, In ZTF, Because CAP1 can collect many contexts (source networks that a user used for access) from RP2 joining in ZTF, by federating with CAP1, RP1 can use the federated contexts about the network which is often used by users.

ZTF allows RPs to share contexts via the CAP beyond identity federations.

\subsubsection{CAP1 Continues to Provide Contexts to RP1 When Changing from IdP2 to IdP1}
Assume that a user uses services provided by the RP1 even though she changes IdP1 to IdP2 for login to RP1 and that RP1 wants to approve access only from the network that users often use.
If IdP1 provides these contexts to RP1, contexts are bound to the identity managed by IdP1 so that after changing to IdP2, RP1 cannot use the contexts having been provided by IdP1.
However, CAP1 can manage contexts independently from the lifecycle of identity managed by IdP1 so that RP1 still uses the contexts provided by CAP1 after changing to IdP2.
Besides, since CAP1 collects contexts from RP2 as well as RP1, CAP1 can provide RP1 federated contexts that only RP1 cannot collect on their own.

ZTF allows CAPs to manage contexts independently from the lifecycle of identity managed by IdP1.

\subsubsection{Minimizing trust of IdP3 in RP2}
Assume that RP2 wants to use contexts to enforce access control based on Zero Trust.
Since RP2 federates with CAP1 and CAP2, RP2 can evaluate the trustworthiness of every access request by using not only federated identity from IdP3 but also federated contexts from CAP1 and CAP2.

RP2 can more minimize implicit trust of IdP3 in ZTF than only in IdF2.

\subsubsection{User Control of Context Sharing}
By using the above use cases, we suggest that ZTF allows RPs to enforce access control based on Zero Trust.
However, since contexts are privacy sensitive information, sharing contexts must be under user control.
In this section, we explain how we implement the user control mechanism.

\begin{figure}[htb]
  \centering
  \includegraphics[width=0.45\textwidth]{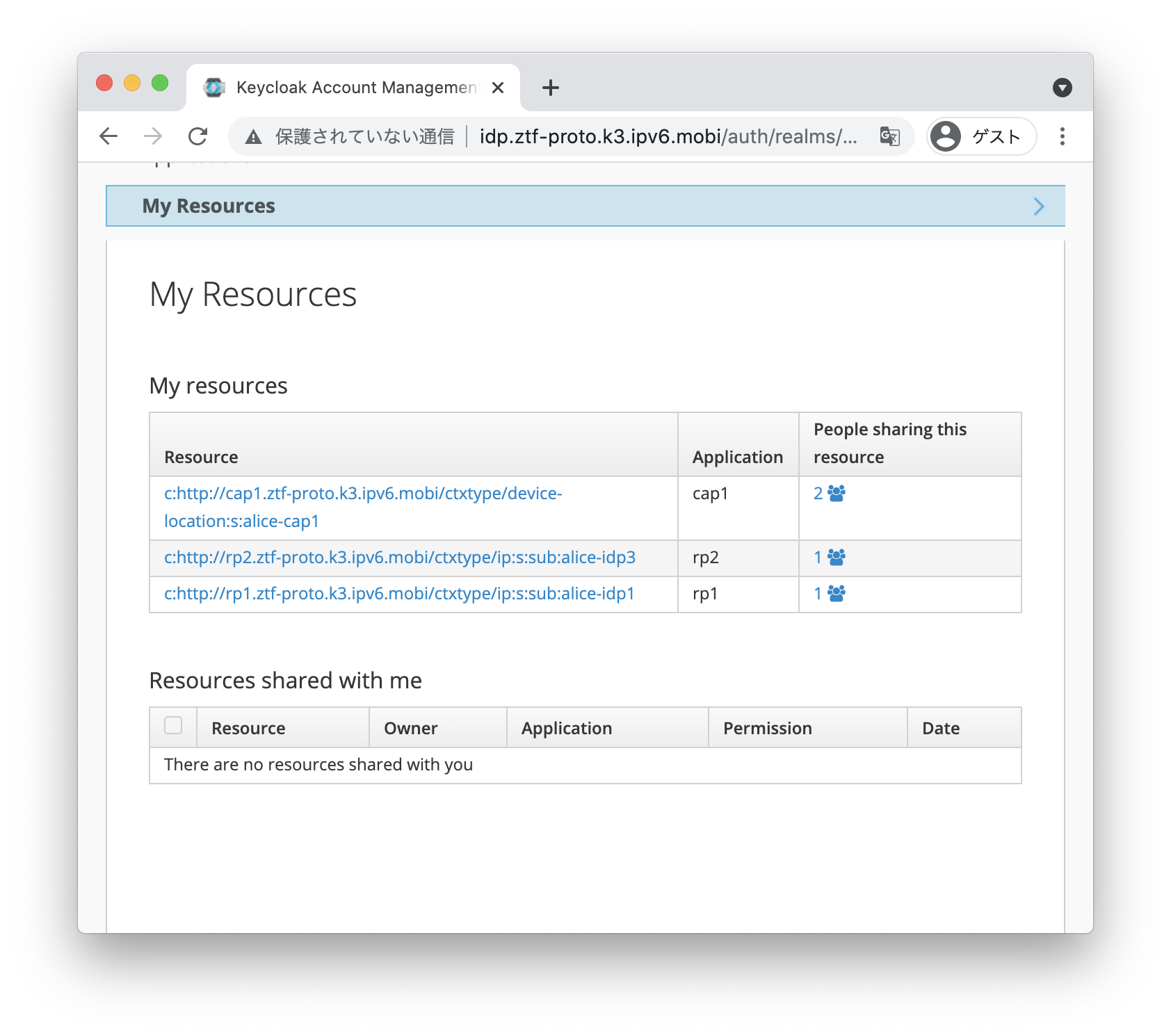}
  \caption{Managed Contexts List in Authorization Server}
  \label{fig:usecases:res-list}
\end{figure}

We implement Authorization Server by using Keycloak, which is open-source software for identity and access management.
Authorization Server manages what kind of contexts CAPs and RPs have and which CAPs and RPs can access what limited contexts.
Fig.\ \ref{fig:usecases:res-list} shows a Web UI in which a user sees what kind of contexts the Authorization Server manages.
For example, this Authorization Server manages contexts of the type (\verb|device-location|) in CAP1.

\begin{figure}[htb]
  \centering
  \includegraphics[width=0.45\textwidth]{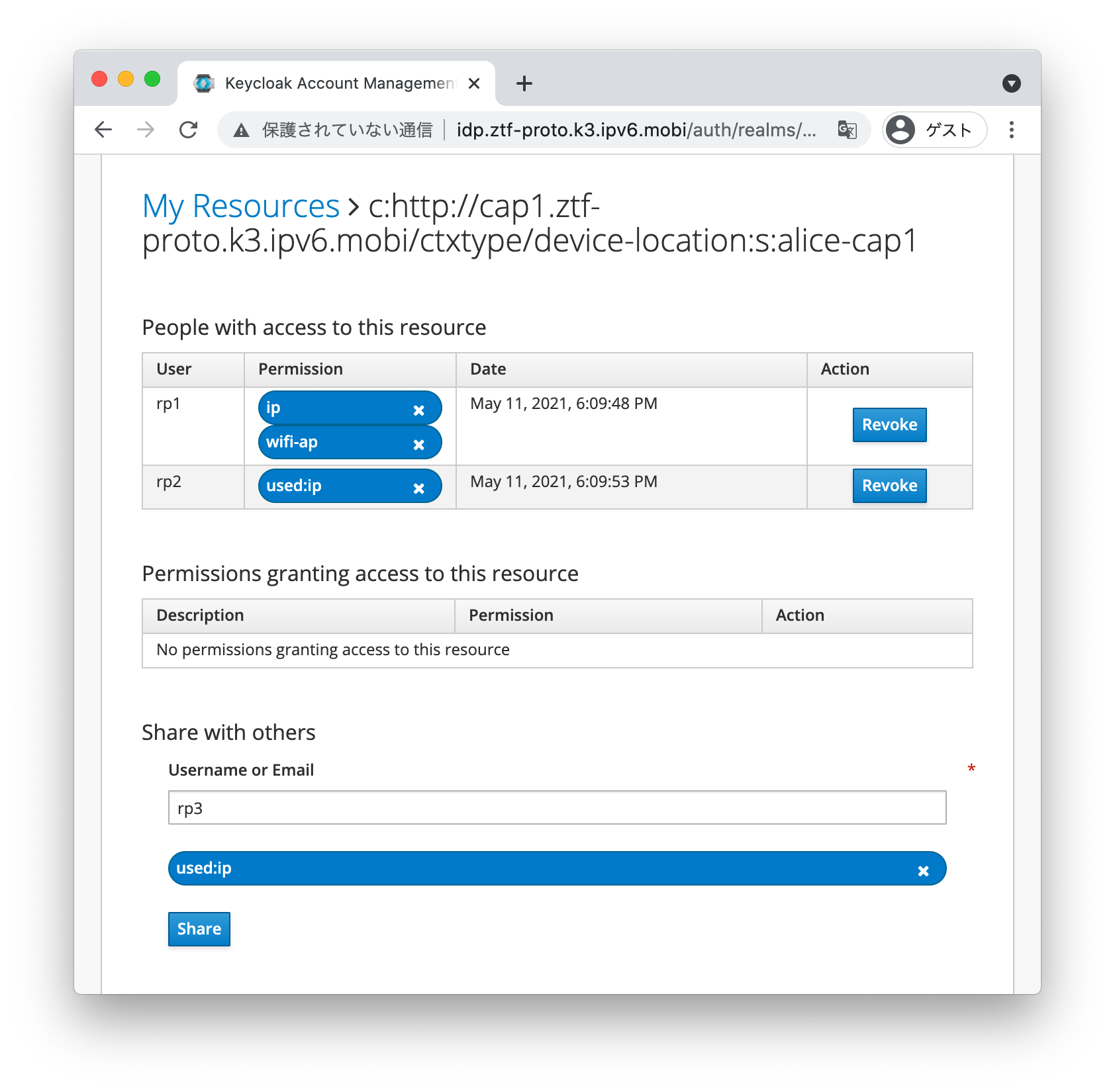}
  \caption{User Managed Contexts in Authorization Server}
  \label{fig:usecases:res-share}
\end{figure}

Fig.\ \ref{fig:usecases:res-share} shows a Web UI in which a user sees and manages which CAPs and RPs can access what limited contexts.
Fig.\ \ref{fig:usecases:res-share} shows that RP1 can access this context type (\verb|device-location|) with limited permission (\verb|ip| and \verb|wifi-ap|) and RP1 can access this context type with limited permission (\verb|used:ip|).
Fig.\ \ref{fig:usecases:res-share} also shows that a user set policy about which RPs can access this context type with which permission by using the below form named ``Share with Others''.
In this example, the user tries to set policy that RP3 can access this context type with limited permission(\verb|used:ip|).

\section{Conclusion}
We proposed the concept of Zero Trust Federation (ZTF), which allows RPs to apply the concept of Zero Trust under identity federations.
In ZTF, we introduced a new entity called Context Attribute Provider that collects and provides contexts across entities in ZTF independently from IdPs and RPs.

We design a mechanism that shares contexts among systems operated by entities joining a ZTF.
When sharing contexts, two protocols are needed: context transport protocol and user authorization protocol.
For the former, we designed a mechanism to transmit and receive contexts using a protocol called CAEP, which is currently being standardized by Shared Signals and Events WG of OpenID Foundation.
For the latter, we designed a user control mechanism with user managed access control protocol called UMA.

We implemented the prototype of ZTF and evaluated the capability of ZTF to allow RPs to enforce access control based on Zero Trust in four use cases.

Future work will address the following problems: 1) standardizing the format and the semantics of contexts in ZTF; 2) operation of the authorization server, e.g., how to write policies and where to deploy them.

\bibliographystyle{ipsjsort-e}
\bibliography{thesis}

\end{document}